# Fastest Distributed Consensus Averaging Problem on Chain of Rhombus Networks

Saber Jafarizadeh
Department of Electrical Engineering
Sharif University of Technology, Azadi Ave, Tehran, Iran
Email: jafarizadeh@ee.sharif.edu

*Abstract*—**Distributed consensus has appeared as one of the most important and primary problems in the context of distributed computation and it has received renewed interest in the field of sensor networks (due to recent advances in wireless communications), where solving fastest distributed consensus averaging problem over networks with different topologies is one of the primary problems in this issue.**

**Here in this work analytical solution for the problem of fastest distributed consensus averaging algorithm over Chain of Rhombus networks is provided, where the solution procedure consists of stratification of associated connectivity graph of the network and semidefinite programming, particularly solving the slackness conditions, where the optimal weights are obtained by inductive comparing of the characteristic polynomials initiated by slackness conditions. Also characteristic polynomial together with its roots corresponding to eigenvalues of weight matrix including *SLEM* of network is determined inductively. Moreover to see the importance of rhombus graphs it is indicated that convergence rate of path network increases by replacing a single node by a rhombus sub graph within the path network.**

## I. INTRODUCTION

Distributed computation in the context of computer science is a well studied field with an extensive body of literature (see, for example, [1] for early work), where some of its applications include distributed agreement, synchronization problems, [2] and load balancing in parallel computers [3,4].

A problem that has received renewed interest recently is distributed consensus averaging algorithms in sensor networks and one of main research directions is the computation of the optimal weights that yields the fastest convergence rate to the asymptotic solution [5, 6, 7] known as Fastest Distributed Consensus averaging Algorithm, which computes iteratively the global average of distributed data in a sensor network by using only local communications. Moreover algorithms for distributed consensus find applications in, e.g., multi-agent distributed coordination and flocking [8, 9, 10, 11], distributed data fusion in sensor networks [12, 13, 6], fastest mixing Markov chain problem [14], clustering [15, 16] gossip algorithms [17, 18], and distributed estimation and detection for decentralized sensor networks [19, 20, 21, 22, 23]. Recently in [24, 25, 26] the author has solved fastest distributed consensus averaging problem analytically for path, complete cored, symmetric and asymmetric star networks and in [27] the author has determined the optimal weights for edges of different types of branches of an arbitrary connected network and has proved that the obtained weights are independent of rest of the network and in [28] fastest distributed consensus averaging problem over one of basic and common types of networks namely, tree network has been studied.

Here in this work, we have provided analytical solution for fastest distributed consensus averaging problem over Chain of Rhombus (CR) networks, by means of stratification and semidefinite programming. Our method in this paper is based on convexity of fastest distributed consensus averaging problem, and inductive comparing of the characteristic polynomials initiated by slackness conditions in order to find the optimal weights. Also characteristic polynomial together with its roots corresponding to eigenvalues of optimal weight matrix including *SLEM* of network is determined inductively and to see the importance of rhombus graphs it is indicated that convergence rate of path network increases by replacing a single node by a rhombus sub graph within the path network. Moreover it is indicated that the obtained optimal weights hold for CR branches independent of rest of the network where these branches are obtained from connecting a CR network to an arbitrary connected network via a bridge.

The organization of the paper is as follows. Section II is an overview of the materials used in the development of the paper, including relevant concepts from distributed consensus averaging algorithm, graph symmetry and semidefinite programming. Section III contains the main results of the paper where CR network and CR branch are introduced together with the corresponding evaluated *SLEM* and obtained optimal weights. Sections IV is devoted to proof of main results of paper for CR network and section V concludes the paper.

## II. PRELIMINARIES

This section introduces the notation used in the paper and reviews relevant concepts from distributed consensus averaging algorithm, graph symmetry and semidefinite programming.

### *A. Distributed Consensus*

We consider a network $\mathcal{N}$ with the associated graph $\mathcal{G} = (\mathcal{V}, \mathcal{E})$ consisting of a set of nodes $\mathcal{V}$ and a set of edges $\mathcal{E}$ where each edge $\{i, j\} \in \mathcal{E}$ is an unordered pair of distinct nodes.

Each node $i$ holds an initial scalar value $x_i(0) \in \mathbf{R}$, and $x^T(0) = (x_1(0), \dots, x_n(0))$ denotes the vector of initial node values on the network. Within the network two nodes can communicate with each other, if and only if they are neighbors.

The main purpose of distributed consensus averaging is to compute the average of the initial values, $(1/n)\sum_{i=1}^{n} x_i(0)$ via a distributed algorithm, in which the nodes only communicate with their neighbors.

In this work, we consider distributed linear iterations, which have the form

$$x_i(t+1) = W_{ii}x_i(t) + \sum_{j \neq i} W_{ij}x_j(t), \quad i = 1, \dots, n$$

where $t = 0,1,2, \dots$ is the discrete time index and $W_{ij}$ is the weight on $x_j$ at node $i$ and the weight matrix have the same sparsity pattern as the adjacency matrix of the network's associated graph or $W_{ij} = 0$ if $\{i, j\} \notin \mathcal{E}$, this iteration can be written in vector form as

$$x(t+1) = Wx(t)$$

The linear iteration (1) implies that $x(t) = W^t x(0)$ for $t = 0,1,2, \dots$ . We want to choose the weight matrix $W$ so that for any initial value $x(0)$, $x(t)$ converges to the average vector $\bar{x} = (\mathbf{1}^T x(0)/n)\mathbf{1} = (\mathbf{1}\mathbf{1}^T/n)x(0)$ i.e.

$$\lim_{t \to \infty} x(t) = \lim_{t \to \infty} W^t x(0) = \frac{\mathbf{1}\mathbf{1}^T}{n} x(0)$$

(Here **1** denotes the column vector with all coefficients one). This is equivalent to the matrix equation

$$\lim_{t \to \infty} W^t = \frac{\mathbf{1}\mathbf{1}^T}{n} \tag{1}$$

Assuming (1) holds, the *convergence factor* can be defined as

$$r(W) = sup \frac{\|x(t+1) - \bar{x}\|_2}{\|x(t) - \bar{x}\|_2}$$

where $\|\cdot\|_2$ denotes the spectral norm, or maximum singular value. The FDC problem in terms of the convergence factor can be expressed as the following optimization problem:

$$\begin{aligned} \min_{W} \quad & r(W) \\ s.t. \quad & \lim_{t\to\infty} W^t = \mathbf{1}\mathbf{1}^T/n, \\ & \forall\{i,j\} \notin \mathcal{E}: W_{ij} = 0 \end{aligned} \qquad (2)$$

where $W$ is the optimization variable, and the network is the problem data.

The FDC problem (2) is closely related to the problem of finding the fastest mixing Markov chain on a graph [14]; the only difference in the two problem formulations is that in the FDC problem, the weights can be (and the optimal ones often are) negative, hence faster convergence could be achieved compared with the fastest mixing Markov chain on the same graph.

In [5] it has been shown that the necessary and sufficient conditions for the matrix equation (1) to hold is that one is a simple eigenvalue of $W$ associated with the eigenvector $\mathbf{1}$, and all other eigenvalues are strictly less that one in magnitude. Moreover in [5] FDC problem has been formulated as the following minimization problem

$$\begin{aligned} \min_{W} \quad & \max(\lambda_2, -\lambda_n) \\ s.t. \quad & W = W^T, W\mathbf{1} = \mathbf{1} \\ & \forall\{i,j\} \notin \mathcal{E}: W_{ij} = 0 \end{aligned}$$

where $1 = \lambda_1 \geq \lambda_2 \geq \cdots \geq \lambda_n \geq -1$ are eigenvalues of $W$ arranged in decreasing order and $\max(\lambda_2, -\lambda_n)$ is the *Second Largest Eigenvalue Modulus* (*SLEM*) of $W$, and the main problem can be derived in the semidefinite programming form as [5]:

$$\begin{aligned} \min_{W} \quad & s \\ s.t. \quad & -sI \preccurlyeq W - \mathbf{1}\mathbf{1}^T/n \preccurlyeq sI \\ & W = W^T, W\mathbf{1} = \mathbf{1} \\ & \forall\{i,j\} \notin \mathcal{E}: W_{ij} = 0 \end{aligned} \qquad (3)$$

We refer to problem (3) as the Fastest Distributed Consensus (FDC) averaging problem.

*B. Symmetry of Graphs*

An automorphism of a graph $\mathcal{G} = (\mathcal{V}, \mathcal{E})$ is a permutation $\sigma$ of $\mathcal{V}$ such that $\{i,j\} \in \mathcal{E}$ if and only if $\{\sigma(i), \sigma(j)\} \in \mathcal{E}$, the set of all such permutations, with composition as the group operation, is called the automorphism group of the graph and denoted by $Aut(\mathcal{G})$. For a vertex $i \in \mathcal{V}$, the set of all images $\sigma(i)$, as $\sigma$ varies through a subgroup $G \subseteq Aut(\mathcal{G})$, is called the orbit of $i$ under the action of $G$. The vertex set $\mathcal{V}$ can be written as disjoint union of distinct orbits. In [29], it has been shown that the weights on the edges within an orbit must be the same.

### *C. Semidefinite Programming*

SDP is a particular type of convex optimization problem [30]. An SDP problem requires minimizing a linear function subject to a linear matrix inequality (LMI) constraint [31]:

$$\begin{aligned} \min \quad & \rho = c^T x, \\ s.t. \quad & F(x) \geq 0 \end{aligned}$$

where $c$ is a given vector, $x^T = (x_1, \dots, x_n)$, and $F(x) = F_0 + \sum_i x_i F_i$, for some fixed Hermitian matrices $F_i$. The inequality sign in $F(x) \geq 0$ means that $F(x)$ is positive semidefinite.

This problem is called the primal problem. Vectors $x$ whose components are the variables of the problem and satisfy the constraint $F(x) \geq 0$ are called primal feasible points, and if they satisfy $F(x) \geq 0$, they are called strictly feasible points. The minimal objective value $c^T x$ is by convention denoted by $\rho^*$ and is called the primal optimal value.

Due to the convexity of the set of feasible points, SDP has a nice duality structure, with the associated dual program being:

$$\begin{aligned} \max \quad & -Tr[F_0 Z] \\ s.t. \quad & Z \geq 0 \\ & Tr[F_i Z] = c_i \end{aligned}$$

Here the variable is the real symmetric (or Hermitian) positive matrix $Z$, and the data $c$, $F_i$ are the same as in the primal problem. Correspondingly, matrix $Z$ satisfying the constraints is called dual feasible (or strictly dual feasible if $Z > 0$). The maximal objective value of $-Tr[F_0 Z]$, i.e. the dual optimal value is denoted by $d^*$.

The objective value of a primal (dual) feasible point is an upper (lower) bound on $\rho^* (d^*)$. The main reason why one is interested in the dual problem is that one can prove that $d^* \leq \rho^*$, and under relatively mild assumptions, we can have $\rho^* = d^*$. If the equality holds, one can prove the following optimality condition on $x$.

A primal feasible $x$ and a dual feasible $Z$ are optimal, which is denoted by $\hat{x}$ and $\hat{Z}$, if and only if

$$F(\hat{x})\hat{Z} = \hat{Z}F(\hat{x}) = 0. \tag{4}$$

This latter condition is called the complementary slackness condition.

In one way or another, numerical methods for solving SDP problems always exploit the inequality $d \leq d^* \leq \rho^* \leq \rho$, where $d$ and $\rho$ are the objective values for any dual feasible point and primal feasible point, respectively. The difference

$$\rho^* - d^* = c^T x + Tr[F_0 Z] = Tr[F(x)Z] \geq 0$$

is called the duality gap. If the equality $d^* = \rho^*$ holds, i.e. the optimal duality gap is zero, and then we say that strong duality holds.

## III. MAIN RESULTS

This section presents the main results of the paper. Here we introduce Chain of Rhombus (CR) network and CR branches with the corresponding evaluated *SLEM* and optimal weights, proofs and more detailed discussion are deferred to section IV.

### *A. Chain of Rhombus (CR) Network*

A rhombus graph of order $n$ is a graph with $n$ nodes isolated from each other where each node is connected to two ending nodes of graph (see Fig. 1. for a rhombus graph of order 4) and a CR graph of order $m$ consists of $m$ rhombus graphs of different orders namely $n_1, \dots, n_m$ which shares their ending nodes together as depicted in Fig.2. for $m = 3, n_1 = 3, n_2 = 2, n_3 = 4$.

In section IV we have proved that in a CR network of order $m$ optimal weights of the edges on $i$-th rhombus equal $1/(n_i + 1)$, where $n_i$ for $i = 1, \dots, m$ is the order of $i$-th rhombus, also *SLEM* of CR network for different values of $m$ and $n_1, \dots, n_m$ is presented in table.1. A special case of CR network is a path network with $2m + 1$ nodes which is a CR network for $n_1 = n_2 = \cdots = n_m = 1$ and considering the results obtained in section IV all of the weights on edges equal $1/2$ which is in total agreement with the results obtained in [25, 1].

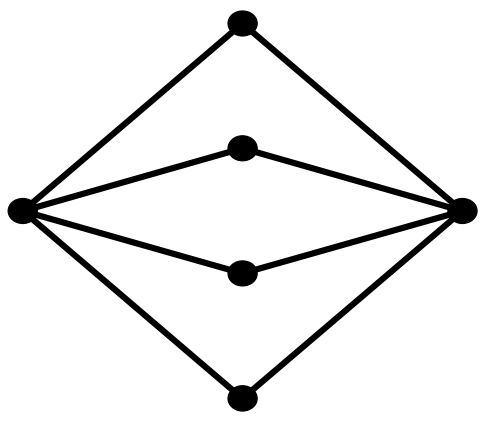

Fig.1. Rhombus graph of order 4.

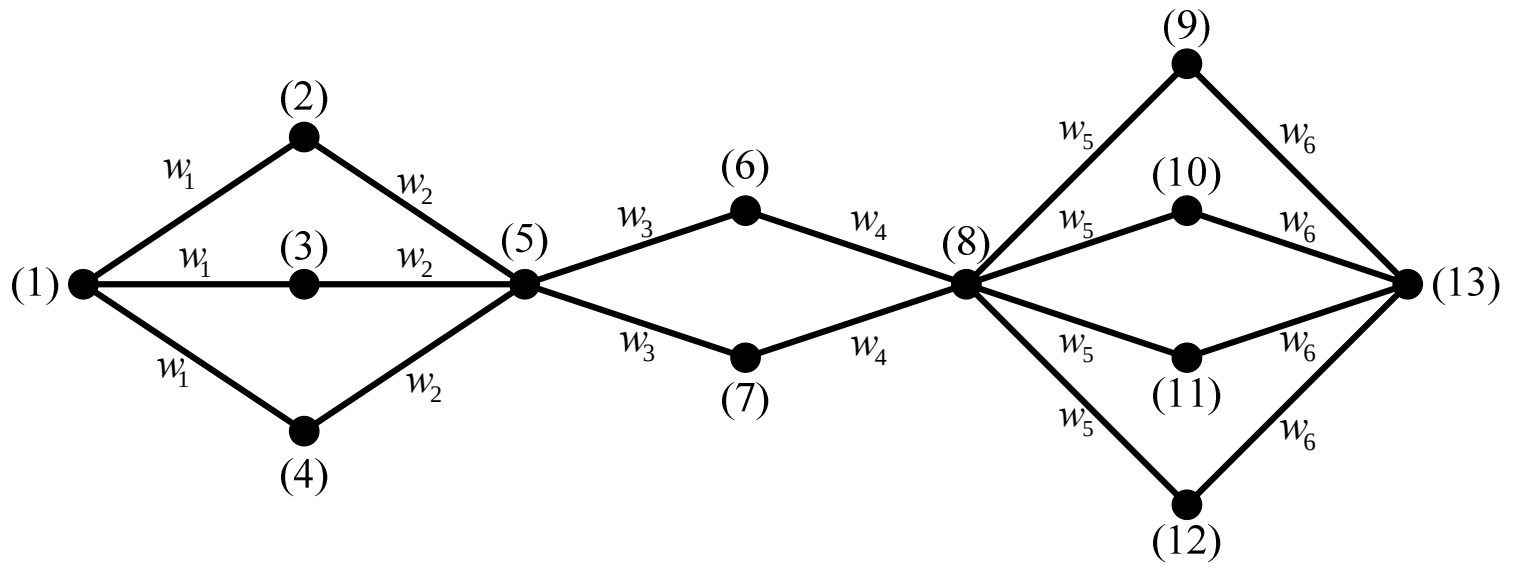

Fig.2. Weighted CR Network for $m = 3, n_1 = 3, n_2 = 2, n_3 = 4$.

| $(m, n_1, \dots, n_m)$ | *SLEM* | $(m, n)$ | *SLEM* | $(m, n)$ | *SLEM* |
|---|---|---|---|---|---|
| $(2,2,2)$ | 0.8047 | $(3,1,1,1)$ | 0.9010 | $(4,1,1,1,1)$ | 0.9397 |
| $(2,2,3)$ | 0.8143 | $(3,1,2,1)$ | 0.8857 | $(4,2,1,1,1)$ | 0.9460 |
| $(2,2,4)$ | 0.8233 | $(3,1,1,2)$ | 0.9091 | $(4,1,2,1,1)$ | 0.9353 |
| $(2,2,5)$ | 0.8306 | $(3,1,3,1)$ | 0.8797 | $(4,1,2,2,1)$ | 0.9289 |
| $(2,2,10)$ | 0.8510 | $(3,1,1,3)$ | 0.9162 | $(4,1,2,1,2)$ | 0.9428 |
| $(2,2,20)$ | 0.8648 | $(3,1,50,1)$ | 0.8669 | $(4,2,1,1,2)$ | 0.9521 |
| $(2,3,3)$ | 0.8257 | $(3,2,2,2)$ | 0.9031 | $(4,3,1,1,1)$ | 0.9507 |
| $(2,3,4)$ | 0.8360 | $(3,2,3,2)$ | 0.8969 | $(4,1,3,1,1)$ | 0.9349 |
| $(2,3,5)$ | 0.8443 | $(3,2,2,3)$ | 0.9116 | $(4,1,3,3,1)$ | 0.9262 |
| $(2,3,10)$ | 0.8671 | $(3,2,4,2)$ | 0.8935 | $(4,1,3,1,3)$ | 0.9491 |
| $(2,5,5)$ | 0.8654 | $(3,2,50,2)$ | 0.8829 | $(4,3,1,1,3)$ | 0.9609 |
| $(2,10,10)$ | 0.9181 | $(3,2,4,3)$ | 0.9027 | $(4,2,2,2,2)$ | 0.9423 |
| $(2,20,20)$ | 0.9548 | $(3,3,3,3)$ | 0.9146 | $(4,3,2,2,2)$ | 0.9476 |
| $(2,50,50)$ | 0.9808 | $(3,3,4,3)$ | 0.9117 | $(4,2,3,2,2)$ | 0.9409 |
| $(2,100,100)$ | 0.9902 | $(3,3,3,4)$ | 0.9214 | $(4,2,3,3,2)$ | 0.9393 |

Table.1. *SLEM* of CR network for different values of $m$ and $n_1, \dots, n_m$.

Defining inner rhombuses as rhombus sub graphs which are connected to other rhombus sub graphs from both sides we can conclude that every CR network of order $m$ has $m-2$ inner rhombuses and as it is obvious from the results depicted in Table. 1. the *SLEM* of CR network decreases with the order of inner rhombuses of network where this happens due to the bottleneck property of inner rhombuses in CR network. Using this property of CR networks, in a path network one can replace a single node by multiple parallel nodes to form an inner rhombus sub graph and decrease SLEM of network which in turn will result in faster convergence rate.

*B. Chain of Rhombus (CR) Branch*

A CR branch consists of a CR network connected to an arbitrary network by a bridge (see Fig.3. for $m = 3, n_1 = 3, n_2 = 2, n_3 = 4$).

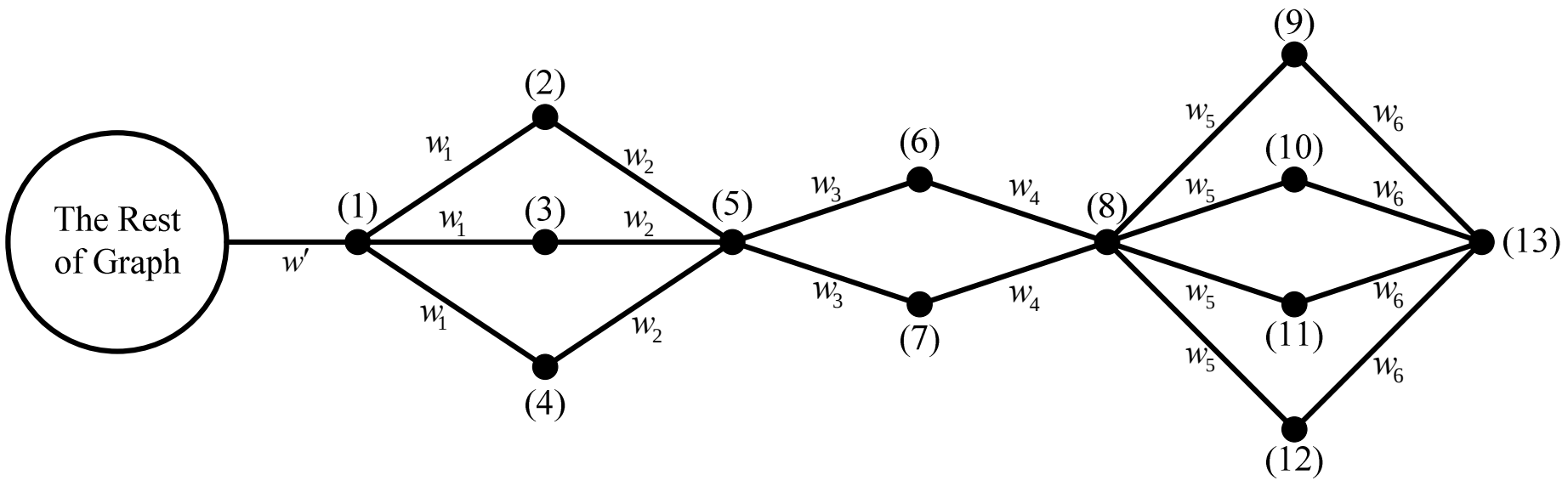

Fig.3. Weighted CR branch for $m=3, n_1=3, n_2=2, n_3=4$.

Using the same inductive procedure as in section IV, we can state that the optimal weights for CR branch are the same as the optimal weights obtained in sections IV for CR network except for the bridge (weighted by $w'$in Fig.3.) which is connecting CR network to the arbitrary network. Specifying optimal weight of bridge and *SLEM* of the network requires comprehensive knowledge about the topology of network.

## IV. Proofs of Main Results

In this section we provide the solution of fastest distributed consensus averaging problem and determine the optimal weights for CR network introduced in section III.

Here we consider a CR network of order $m$ with the undirected associated connectivity graph $\mathcal{G}=(\mathcal{V},\mathcal{E})$ consisting of $|\mathcal{V}|=m+1+\sum_{i=1}^{m} n_i$ nodes and $|\mathcal{E}|=2\sum_{i=1}^{m} n_i$ edges, where the set of nodes is denoted by $\mathcal{V}=\{(1),(2),\dots,(|\mathcal{V}|)\}$ (see Fig.2 for $m=3, n_1=3, n_2=2, n_3=4$).

Automorphism of CR network is $S_{n_i}$ permutation of nodes on $i$-th rhombus for $i=1,\dots,m$, hence according to subsection II-B it has $2m+1$ class of edge orbits and it suffices to consider just $2m$ weights $w_1, w_2, \dots, w_{2m}$ (as labeled in Fig. 2. for $m=3, n_1=3, n_2=2, n_3=4$), and consequently the weight matrix for the network can be defined as

$$W_{i,j}=\begin{cases} w_{2\mu-1} \ \ \textbf{for} \ \ i=\mu+N_{\mu-1}, \ \ j=i+1,\dots,i+n_\mu, \ \ \mu=1,\dots,m \\ w_{2\mu-1} \ \ \textbf{for} \ \ j=\mu+N_{\mu-1}, \ \ i=j+1,\dots,j+n_\mu, \ \ \mu=1,\dots,m \\ w_{2\mu} \ \ \textbf{for} \ \ i=\mu+1+N_\mu, \ \ j=i-1,\dots,i-n_\mu, \ \ \mu=1,\dots,m \\ w_{2\mu} \ \ \textbf{for} \ \ j=\mu+1+N_\mu, \ \ i=j-1,\dots,i-n_\mu, \ \ \mu=1,\dots,m \\ 1-w_{2\mu-1}-w_{2\mu} \ \ \textbf{for} \ \ i=j=N_{\mu-1}+\mu+1,\dots,\mu+n_\mu, \ \ \mu=1,\dots,m \\ 1-n_1 w_1 \ \ \textbf{for} \ \ i=j=1 \\ 1-n_\mu w_{2\mu}-n_{\mu+1}w_{2\mu+1} \ \ \textbf{for} \ \ i=j=N_\mu+\mu+1, \ \ \mu=1,\dots,m-1 \\ 1-n_m w_{2m} \ \ \textbf{for} \ \ i=j=N_m+m+1 \end{cases}$$

where $N_\mu$ equals $\sum_{\rho=1}^{\mu} n_\rho$ for $\mu=1,\dots,m$ and zero for $\mu=0$. We associate with the node $(i)$, the $|\mathcal{V}|\times 1$ column vector $e_i \in \mathbf{R}^{|\mathcal{V}|}$ with 1 in the $i$-th position, and zero elsewhere. Introducing the new basis

$$\varphi_{2i,\mu} = \frac{1}{\sqrt{n_i}} \sum_{k=0}^{n_i-1} \omega_i^{k\mu_i} e_{N_{i-1}+k+2} \quad \textbf{for} \;\; i = 1, \dots, m, \;\; \mu_i = 0, \dots, n_i - 1$$

with $\omega_i = e^{j\frac{2\pi}{n_i}}$, for $i = 1, \dots, m$ the weight matrix $W$ for CR network in the new basis takes the block diagonal form with diagonal blocks $W_0, W_1, W_2, \dots, W_m$ defined as:

$$W_0 = \begin{bmatrix} 1-n_1w_1 & \sqrt{n_1}w_1 & 0 & & & & \\ \sqrt{n_1}w_1 & 1-w_1-w_2 & \sqrt{n_1}w_2 & 0 & & & \\ 0 & \sqrt{n_1}w_2 & 1-n_1w_2-n_2w_3 & \sqrt{n_2}w_3 & 0 & & \\ & 0 & \sqrt{n_2}w_3 & 1-w_3-w_4 & \sqrt{n_2}w_4 & \ddots & \\ & & 0 & \sqrt{n_2}w_4 & \ddots & \ddots & 0 \\ & & & \ddots & \ddots & 1-w_{2m-1}-w_{2m} & \sqrt{n_m}w_{2m} \\ & & & & 0 & \sqrt{n_m}w_{2m} & 1-n_mw_{2m} \end{bmatrix} \tag{5-a}$$

$$W_i = \begin{bmatrix} 1-w_1-w_2 & 0 & \cdots & 0 \\ 0 & 1-w_3-w_4 & \ddots & \vdots \\ \vdots & \ddots & \ddots & 0 \\ 0 & \cdots & 0 & 1-w_{2m-1}-w_{2m} \end{bmatrix} \quad \textbf{for} \;\; i = 1, \dots, m$$

Since $W_i$ is a diagonal matrix for $i = 1, \dots, m$ with the same diagonal elements at each matrix then we can define the diagonal matrix $W_0'$ as follows

$$W_0' = \begin{bmatrix} 1-w_1-w_2 & 0 & & \\ 0 & 1-w_3-w_4 & \ddots & \\ & \ddots & \ddots & 0 \\ & & 0 & 1-w_{2m-1}-w_{2m} \end{bmatrix} \tag{5-b}$$

where the diagonal elements of $W_0'$ are singular values of the weight matrix which are the eigenvalues of $W$ as well

Considering the fact that each element of $W_0'$ is also an element of $W_0$ and using the *Cauchy interlacing theorem*

*Theorem 1* (*Cauchy Interlacing Theorem*) [32]:

Let $A$ and $B$ be $n \times n$ and $m \times m$ matrices, where $m \leq n$, $B$ is called a compression of $A$ if there exists an orthogonal projection $P$ onto a subspace of dimension $m$ such that $PAP = B$. The Cauchy interlacing theorem states that If the eigenvalues of $A$ are $\lambda_1(A) \leq \cdots \leq \lambda_n(A)$, and those of $B$ are $\lambda_1(B) \leq \cdots \leq \lambda_m(B)$, then for all $j$,

$$\lambda_j(A) \leq \lambda_j(B) \leq \lambda_{n-m+j}(A)$$

Notice that, when $n - m = 1$, we have

$$\lambda_j(A) \leq \lambda_j(B) \leq \lambda_{j+1}(A)$$

we can state the following corollary for the elements of $W_0'$ ,

In the case of $m = 1$ , after stratification the weight matrix $W_0$ does not include diagonal entries $1 - n_i w_{2i} - n_{i+1} w_{2i+1}$ **for** $i = 1, \dots, m-1$ and consequently Cauchy interlacing theorem will not be true thus the followings are true for $m \geq 2$.

*Corollary 1*,

For $W_0$ and $W_0'$ given in (5), theorem 1 implies the following relations between the elements of $W_0'$ and eigenvalues of $W_0$

$$\lambda_{|\mathcal{V}|}(W) = \lambda_{2m+1}(W_0) \leq 1 - w_{2i-1} - w_{2i} \leq \lambda_1(W_0) = 1 \quad \textbf{for} \quad i = 1, \dots, m \tag{6}$$

It is obvious from above relations that $\lambda_{|\mathcal{V}|}(W)$ is amongst the eigenvalues of $W_0$ and $\lambda_2(W)$ can be amongst the eigenvalues of both $W_0$ and $W_0'$.

Based on corollary 1 and subsection II-A, one can express FDC problem for CR network in the form of semidefinite programming as:

$$\begin{aligned} & \min \quad s \\ s.t. \quad & -sI \leq W_0 - \boldsymbol{v}\boldsymbol{v}^T \leq sI \\ & W_0' \leq sI \end{aligned} \tag{7}$$

where $\boldsymbol{v}$ is a $(2m+1) \times 1$ column vector defined as:

$$\boldsymbol{v}(i) = \begin{cases} 1 \quad \textbf{for} \quad i = 2\mu + 1, & \mu = 0, \dots, m \\ \sqrt{n_\mu} \quad \textbf{for} \quad i = 2\mu, & \mu = 1, \dots m \end{cases}$$

which is eigenvector of $W_0$ corresponding to the eigenvalue one. The matrices $W_0$ and $W_0'$ can be written as

$$W_0 = I_{2m+1} - \sum_{i=1}^{2m} w_i \, \boldsymbol{\alpha}_i \boldsymbol{\alpha}_i^T \tag{8-a}$$

$$W_0' = I_m - \sum_{i=1}^{m} (w_{2i-1} + w_{2i}) \boldsymbol{\beta}_i \boldsymbol{\beta}_i^T \tag{8-b}$$

where $\boldsymbol{\alpha}_i$ and $\boldsymbol{\beta}_i$ are $(2m+1) \times 1$ and $m \times 1$ column vectors, respectively defined as:

$$\boldsymbol{\alpha}_{2i-1}(j) = \begin{cases} \sqrt{n_i} & j = 2i - 1 \\ -1 & j = 2i \\ 0 & Otherwise \end{cases} \quad \textbf{for} \quad i = 1, \dots, m,$$

$$\boldsymbol{\alpha}_{2i}(j) = \begin{cases} 1 & j = 2i \\ -\sqrt{n_i} & j = 2i + 1 \\ 0 & Otherwise \end{cases} \quad \textbf{for} \quad i = 1, \dots, m,$$

$$\boldsymbol{\beta}_i(j) = \begin{cases} 1 & j = i \\ 0 & Otherwise \end{cases} \quad \textbf{for} \quad i = 1, \dots, m.$$

In order to formulate problem (7) in the form of standard semidefinite programming described in section II-C, we define $F_i, c_i$ and $x$ as below:

$$F_0 = \begin{bmatrix} -I_{2m+1} + \boldsymbol{v}\boldsymbol{v}^T & 0 & 0 \\ 0 & I_{2m+1} - \boldsymbol{v}\boldsymbol{v}^T & 0 \\ 0 & 0 & -I_m \end{bmatrix}$$

$$F_i = \begin{bmatrix} \boldsymbol{\alpha}_i\boldsymbol{\alpha}_i^T & 0 & 0 \\ 0 & -\boldsymbol{\alpha}_i\boldsymbol{\alpha}_i^T & 0 \\ 0 & 0 & \boldsymbol{\beta}_{\lfloor\frac{i+1}{2}\rfloor}\boldsymbol{\beta}_{\lfloor\frac{i+1}{2}\rfloor}^T \end{bmatrix} \quad \mathbf{for} \quad i = 1, \ldots, 2m,$$

$$F_{2m+1} = I_{5m+2}$$

$$c_i = 0, \;\; i = 1, \ldots 2m, \quad c_{2m+1} = 1$$

$$x^T = [w_1, w_2, \ldots, w_{2m}, s]$$

In the dual case we choose the dual variable $Z \geq 0$ as

$$Z = \begin{bmatrix} z_1 \\ z_2 \\ z_3 \end{bmatrix} \cdot [z_1^T \quad z_2^T \quad z_3^T] \tag{9}$$

where $z_1, z_2$ and $z_3$, are$(2m+1) \times 1$ and $m \times 1$ column vectors, respectively. Obviously (9) choice of $Z$ implies that it is positive definite.

From the complementary slackness condition (4) we have

$$(sI - W_0 + \boldsymbol{v}\boldsymbol{v}^T)z_1 = 0 \tag{10-a}$$

$$(sI + W_0 - \boldsymbol{v}\boldsymbol{v}^T)z_2 = 0 \tag{10-b}$$

$$(sI - W_0')z_3 = 0 \tag{10-c}$$

Multiplying both sides of equations (10-a) and (10-b) by $\boldsymbol{v}\boldsymbol{v}^T$ we have $s(\boldsymbol{v}\boldsymbol{v}^T z_1) = 0$ and $s(\boldsymbol{v}\boldsymbol{v}^T z_2) = 0$ respectively which implies that

$$\boldsymbol{v}^T z_1 = 0 \tag{11-a}$$

$$\boldsymbol{v}^T z_2 = 0 \tag{11-b}$$

Using the constraints $Tr[F_i Z] = c_i$ we have

$$z_1^T z_1 + z_2^T z_2 + z_3^T z_3 = 1 \tag{12-a}$$

$$(\boldsymbol{\alpha}_i^T z_1)^2 + \left(\boldsymbol{\beta}_{\lfloor\frac{i+1}{2}\rfloor}^T z_3\right)^2 = (\boldsymbol{\alpha}_i^T z_2)^2 \quad \mathbf{for} \quad i = 1, \ldots, 2m \tag{12-b}$$

To have the strong duality we set $c^T x + Tr[F_0 Z] = 0$, hence we have

$$z_1^T z_1 - z_2^T z_2 + z_3^T z_3 = s$$

Considering the linear independence of $\boldsymbol{\alpha}_i$ for $i = 1, \dots, 2m$, and $\boldsymbol{\beta}_i$ for $i = 1, \dots, m$, we can expand $z_1, z_2$ and $z_3$ in terms of $\boldsymbol{\alpha}_i$ and $\boldsymbol{\beta}_i$ as

$$z_1 = \sum_{i=1}^{2m} a_i \boldsymbol{\alpha}_i \tag{13-a}$$

$$z_2 = \sum_{i=1}^{2m} a_i' \boldsymbol{\alpha}_i \tag{13-b}$$

$$z_3 = \sum_{i=1}^{m} b_i \boldsymbol{\beta}_i \tag{13-c}$$

with the coordinates $a_i, a_i'$ for $i = 1, \dots, 2m$ and $b_i$, $i = 1, \dots, m$ to be determined.

Using (8) and the expansions (13), while considering (11), by comparing the coefficients of $\boldsymbol{\alpha}_i$ for $i = 1, \dots, 2m$ and $\boldsymbol{\beta}_i$ for $i = 1, \dots, m$ in the slackness conditions (10), we have

$$(-s+1)a_i = w_i \boldsymbol{\alpha}_i^T z_1 \quad \textbf{for} \quad i = 1, \dots, 2m \tag{14-a}$$

$$(s+1)a_i' = w_i \boldsymbol{\alpha}_i^T z_2 \quad \textbf{for} \quad i = 1, \dots, 2m \tag{14-b}$$

$$(-s+1)b_i = (w_{2i-1} + w_{2i}) \boldsymbol{\beta}_i^T z_3 \quad \textbf{for} \quad i = 1, \dots, m \tag{14-c}$$

Considering (12-c), we obtain

$$\left(\frac{(-s+1)a_i}{w_i}\right)^2 + \left(\frac{(-s+1)b_{\lfloor \frac{i+1}{2} \rfloor}}{\left(w_{2\lfloor \frac{i+1}{2} \rfloor -1} + w_{2\lfloor \frac{i+1}{2} \rfloor}\right)}\right)^2 = \left(\frac{(s+1)a_i'}{w_i}\right)^2 \quad for \quad i = 1, \dots, 2m \tag{15}$$

for $\boldsymbol{\alpha}_i^T z_1, \boldsymbol{\alpha}_i^T z_2,\ i = 1, \dots 2m$ and $\boldsymbol{\beta}_i^T z_3\ \ i = 1, \dots m$, we have

$$\boldsymbol{\alpha}_i^T z_1 = \sum_{j=1}^{n} a_j G_{i,j} \tag{16-a}$$

$$\boldsymbol{\alpha}_i^T z_2 = \sum_{j=1}^{n} a_j' G_{i,j} \tag{16-b}$$

$$\boldsymbol{\beta}_i^T z_3 = b_i \quad \textbf{for} \quad i = 1, \dots m \tag{16-c}$$

where $G$ is the gram matrix, defined as

$$G_{i,j} = \boldsymbol{\alpha}_i^T \boldsymbol{\alpha}_j$$

or equivalently

$$G = \begin{bmatrix} n_1+1 & -1 & 0 & & & & \cdots & 0 \\ -1 & n_1+1 & -\sqrt{n_1 n_2} & 0 & & & & \vdots \\ 0 & -\sqrt{n_1 n_2} & n_2+1 & -1 & 0 & & & \\ & 0 & -1 & n_2+1 & -\sqrt{n_2 n_3} & 0 & & \\ & & 0 & -\sqrt{n_2 n_3} & n_3+1 & -1 & \ddots & \\ & & & 0 & -1 & n_3+1 & \ddots & 0 \\ \vdots & & & & \ddots & \ddots & \ddots & -1 \\ 0 & \cdots & & & & 0 & -1 & n_m+1 \end{bmatrix}$$

Substituting (16) in (14) we have

$$(-s+1-(n_1+1)w_1)a_1 = -w_1 a_2 \quad \text{(17-a)}$$

$$(-s+1-(n_i+1)w_{2i-1})a_{2i-1} = -w_{2i-1}\left(\sqrt{n_{i-1}n_i}\,a_{2i-2} + a_{2i}\right) \quad \textbf{for} \quad i = 2, \dots, m \quad \text{(17-b)}$$

$$(-s+1-(n_i+1)w_{2i})a_{2i} = -w_{2i}\left(a_{2i-1} + \sqrt{n_i n_{i+1}}\,a_{2i+1}\right) \quad \textbf{for} \quad i = 1, \dots, m-1 \quad \text{(17-c)}$$

$$(-s+1-(n_m+1)w_{2m})a_{2m} = -w_{2m}a_{2m-1} \quad \text{(17-d)}$$

and

$$\left(s+1-w_1(n_1+1)\right)a'_1 = -w_1 a'_2 \quad \text{(18-a)}$$

$$(s+1-(n_i+1)w_{2i-1})a'_{2i-1} = -w_{2i-1}\left(\sqrt{n_{i-1}n_i}\,a'_{2i-2} + a'_{2i}\right) \quad \textbf{for} \quad i = 2, \dots, m \quad \text{(18-b)}$$

$$(s+1-(n_i+1)w_{2i})a'_{2i} = -w_{2i}\left(a'_{2i-1} + \sqrt{n_i n_{i+1}}\,a'_{2i+1}\right) \quad \textbf{for} \quad i = 1, \dots, m-1 \quad \text{(18-c)}$$

$$(s+1-(n_m+1)w_{2m})a'_{2m} = -w_{2m}a'_{2m-1} \quad \text{(18-d)}$$

and

$$(-s+1-w_{2i-1}-w_{2i})b_i = 0 \quad \textbf{for} \quad i = 1, \dots m \quad (19)$$

By assuming that *SLEM* does not equal to $1 - w_{2i-1} - w_{2i}$ for $i = 1, \dots m$ from (19) we conclude that

$$b_i = 0 \quad \textbf{for} \quad i = 1, \dots m \quad (20)$$

and from recursive equations (17) and (18) and relations (15) we obtain the optimal weights as

$$w_{2i-1} = w_{2i} = \frac{1}{n_i + 1} \quad \textbf{for} \quad i = 1, \dots m \quad (21)$$

where the detailed proof is provided in appendix A. Substituting the weights (21) in $W_0$ given in (5-a) we can deduce that

$$1 - n_i w_{2i} - n_{i+1} w_{2i+1} = -(1 - w_{2i-1} - w_{2i}) \ \ \textbf{for} \ \ i = 1, \dots, m-1 \tag{22}$$

and by applying theorem 1 for diagonal entries of $W_0$, namely $1 - n_i w_{2i} - n_{i+1} w_{2i+1}$ for $i = 1, \dots, m-1$ and considering (22) we have

$$-1 = -\lambda_1(W_0) \le 1 - w_{2i-1} - w_{2i} \le -\lambda_{2m+1}(W_0) = -\lambda_{|\mathcal{V}|}(W) \ \ \textbf{for} \ \ i = 1, \dots, m \tag{23}$$

where by considering the fact that for (21) choice of weights $|\lambda_{|\mathcal{V}|}(W)| = |\lambda_{2m+1}(W_0)| = |\lambda_2(W_0)| = |\lambda_2(W)| = SLEM$ we can conclude that

$$|1 - w_{2i-1} - w_{2i}| \le SLEM \ \ \textbf{for} \ \ i = 1, \dots, m$$

which in turn supports above assumption that *SLEM* does not equal to $1 - w_{2i-1} - w_{2i}$ for $i = 1, \dots m$.

In [33] it is shown that the roots of three term recursive equations remain simple until all of their offdiagonal entries are nonzero and in view of the fact that for positive values of $w_i$ and $n_i$ for $i = 1, \dots, 2m$ all offdiagonal entries of (17) and (18) are nonzero then all roots of equations (17) and (18) or equivalently roots of (32) are simple for positive values of $w_i$ and $n_i$ for $i = 1, \dots, 2m$ which in turn means that ordering of the roots remains the same for positive values of $w_i$ and $n_i$ for $i = 1, \dots, 2m$. In the case of $n_1 = n_2 = \cdots = n_{2m} = 1$ CR network reduces to a path network with $2m + 1$ nodes where in [25] it is shown that all roots of (32) or eigenvalues of $W_0$ as provided in (5-a) for path network are simple and strictly less than one in magnitude except the first eigenvalue of $W_0$ which is equal to one thus due to constant ordering of roots of (32) all roots of (32) are simple and one is the largest root of (32) for all positive values of $w_i$ and $n_i$ for $i = 1, \dots, 2m$ on the other hand for the optimum weights (21), $|\lambda_{|\mathcal{V}|}(W)| = |\lambda_{2m+1}(W_0)| = |\lambda_2(W_0)| = |\lambda_2(W)| < 1$ (since (32) is an even polynomial) hence for the optimum weights (21) all eigenvalues of $W$ are simple and strictly less than one in magnitude ( except the eigenvalue one associated with the eigenvector **1** ) which is the necessary and sufficient conditions for the convergence of weight matrix.

## V. CONCLUSION

Fastest Distributed Consensus averaging Algorithm in sensor networks has received renewed interest recently, but Most of the methods proposed so far usually avoid the direct computation of optimal weights and deal with the Fastest Distributed Consensus problem by numerical convex optimization methods.

Here in this work, we have analytically solved fastest distributed consensus averaging problem for CR network, by means of stratification and semidefinite programming. Our approach in this paper is based on convexity of fastest distributed consensus averaging problem, and inductive comparing of the characteristic polynomials initiated by slackness conditions in order to find the optimal weights. Also it has been indicated that the obtained optimal weights hold for CR branch independent of the rest of network.

As it is shown in this work in a path network by replacing a single node by an inner rhombus sub graph, convergence rate of path network increases due to bottleneck property of rhombus network which is a great advantage for rhombus network over other networks. In this paper bottleneck property has been studied only over path network but we believe that bottleneck property of rhombus network along with the method used in this paper is powerful and lucid enough to be extended to networks with more general topologies which is the object of our future investigations.

## APPENDIX A

(Determination of optimal weights of CR Network)

In this section the optimal weights for CR network is determined by taking (20) into consideration.

Considering (20) relations (15) reduces to the following

$$\left(\frac{(-s+1)a_i}{w_i}\right)^2 = \left(\frac{(s+1)a'_i}{w_i}\right)^2 \quad \textbf{for} \quad i = 1, \dots, 2m$$

where it can be concluded that

$$\frac{a_i^2}{a_j^2} = \frac{a_i'^2}{a_j'^2} \quad \textbf{for} \quad \forall i, j = 1, \dots, 2m \tag{24}$$

Now we can determine the optimal weights in an inductive manner as follows:

In the first stage, from comparing equations (17-a) and (18-a) and considering the relation (24), we can conclude that

$$(-s + 1 - (n_1 + 1)w_1)^2 = (s + 1 - (n_1 + 1)w_1)^2$$

which results in $w_1 = 1/(n_1 + 1)$ and $s = 0$, where the latter is not acceptable. Substituting $w_1 = 1/(n_1 + 1)$ in (17-a) and (18-a), we have

$$(n_1 + 1)sa_1 = a_2$$

$$-(n_1 + 1)sa'_1 = a'_2$$

Continuing the above procedure inductively, up to $k - 1$ stages, and assuming

$$a_j = f_j(s)a_1, \qquad 1 < \forall j \le k$$

and

$$a'_j = f_j(-s)a'_1 \qquad 1 < \forall j \le k$$

where $f_j(s)$ is an even polynomial in terms of $s$ for odd values of $j$ and an odd polynomial in terms of $s$ for even values of $j$ for the $k$-th stage, by comparing equations (17-b) and (18-b) for odd values of $k$ and (17-c) and (18-c) for even values of $k$, we get the following equations

$$\left((-s+1-(n_i+1)w_{2i-1})f_{2i-1}(s)+w_{2i-1}\sqrt{n_{i-1}n_i}f_{2i-2}(s)\right)a_1=-w_{2i-1}a_{2i} \tag{25-a}$$

$$\left((s+1-(n_i+1)w_{2i-1})f_{2i-1}(-s)+w_{2i-1}\sqrt{n_{i-1}n_i}f_{2i-2}(-s)\right)a'_1=-w_{2i-1}a'_{2i} \tag{25-b}$$

and

$$\left((-s+1-(n_i+1)w_{2i})f_{2i}(s)+w_{2i}f_{2i-1}(s)\right)a_1=-w_{2i}\sqrt{n_i n_{i+1}}a_{2i+1} \tag{26-a}$$

$$\left((s+1-(n_i+1)w_{2i})f_{2i}(-s)+w_{2i}f_{2i-1}(-s)\right)a'_1=-w_{2i}\sqrt{n_i n_{i+1}}a'_{2i+1} \tag{26-b}$$

where (25) and (26) stands for odd and even values of $k$ respectively and considering relation (24) we can conclude that

$$\left((-s+1-(n_i+1)w_{2i-1})f_{2i-1}(s)+w_{2i-1}\sqrt{n_{i-1}n_i}f_{2i-2}(s)\right)^2=\left((s+1-(n_i+1)w_{2i-1})f_{2i-1}(-s)+w_{2i-1}\sqrt{n_{i-1}n_i}f_{2i-2}(-s)\right)^2$$

and

$$\left((-s+1-(n_i+1)w_{2i})f_{2i}(s)+w_{2i}f_{2i-1}(s)\right)^2=\left((s+1-(n_i+1)w_{2i})f_{2i}(-s)+w_{2i}f_{2i-1}(-s)\right)^2$$

which for odd and even values of $k$ results in

$$w_{2i-1}=\frac{1}{n_i+1}$$

$$w_{2i}=\frac{1}{n_i+1}$$

respectively or equivalently

$$w_k=\frac{1}{n_{\lfloor\frac{k+1}{2}\rfloor}+1} \tag{27}$$

Substituting $w_k=1/\left(n_{\lfloor\frac{k+1}{2}\rfloor}+1\right)$ in (25) and (26) for odd and even values of $k$ respectively, we have

$$a_{2i}=\left((n_i+1)sf_{2i-1}(s)-\sqrt{n_{i-1}n_i}f_{2i-2}(s)\right)a_1 \tag{28-a}$$

$$a'_{2i}=\left(-(n_i+1)sf_{2i-1}(-s)-\sqrt{n_{i-1}n_i}f_{2i-2}(-s)\right)a'_1 \tag{28-b}$$

and

$$\frac{1}{\sqrt{n_i n_{i+1}}}\big((n_i+1)sf_{2i}(s)-f_{2i-1}(s)\big)a_1=a_{2i+1} \quad (29\text{-a})$$

$$\frac{1}{\sqrt{n_i n_{i+1}}}\big(-(n_i+1)sf_{2i}(-s)-f_{2i-1}(-s)\big)a'_1=a'_{2i+1} \quad (29\text{-b})$$

where (27), (28) and (29) hold true for $k=1,\dots,2m-1$, $i=2,\dots,m-1$ and $i=1,\dots,m-1$ and in the $2m$-th stage, from equations (17-d) and (18-d) and using relations (24) and (29), we can conclude that

$$w_{2m}=\frac{1}{n_m+1} \quad (30)$$

$$a_{2m}=\Big((n_m+1)sf_{2m-1}(s)-\sqrt{n_{m-1}n_m}f_{2m-2}(s)\Big)\,a_1 \quad (31\text{-a})$$

$$a'_{2m}=\Big(-(n_m+1)sf_{2m-1}(-s)-\sqrt{n_{m-1}n_m}f_{2m-2}(-s)\Big)\,a'_1 \quad (31\text{-b})$$

and $s$ has to satisfy following equation

$$((n_m+1)^2s^2-1)f_{2m-1}(s)=s(n_m+1)\sqrt{n_{m-1}n_m}f_{2m-2}(s) \quad (32)$$

which is obtained from comparing equations (31-a) and (17-d). Equation (32) is an even polynomial for all values of $m$ where it has been computed and provided below for $m=2$ and $m=3$ as

$$(n_2+1)^2(n_1+1)^2s^4-((n_2+1)^2+n_1n_2(n_1+1)(n_2+1)+(n_1+1)^2)s^2+1=0$$

and

$$((n_3+1)^2s^2-1)\times((n_2+1)^2s^2-1)\times((n_1+1)^2s^2-1)+n_1n_2(n_1+1)(n_2+1)s^2\times(1-(n_3+1)^2s^2)$$
$$+n_2n_3(n_3+1)(n_2+1)s^2\times(1-(n_1+1)^2s^2)+n_1n_2n_2n_3(n_1+1)(n_3+1)s^2=0$$

respectively.

## REFFERENCES